\documentclass[11pt]{article}

\usepackage{graphicx,url}
\usepackage{enumerate}

\usepackage{xr}

\def\tends{\rightarrow}

\def\loeq{\leq_{L}}
\def\loe{<_{L}}

\newtheorem{Theorem}{Theorem}

\newtheorem{Proposition}{Proposition}

\newcommand{\Appendix}
{
\def\thesection{Appendix}
\def\thesubsection{A.\arabic{subsection}}
}

\usepackage{amssymb}
\usepackage{amsmath}
\usepackage{mathrsfs}
\usepackage{latexsym}
\usepackage{epsfig}

\date{}

\title{ Constrained Parameterization of Reduced Rank and Co-integrated Vector Autoregression}
\author{
 Anindya Roy\footnote{\baselineskip=10pt Center for Statistical Research and Methodology, U.S. Census Bureau, 4600 Silver Hill Road, Washington, D.C. 20233-9100, anindya.roy@census.gov}  and Tucker S. McElroy\footnote{\baselineskip=10pt Research and Methodology Directorate, U.S. Census Bureau, 4600 Silver Hill Road, Washington, D.C. 20233-9100, tucker.s.mcelroy@census.gov}  \footnote{\baselineskip=10pt This report is released to inform interested parties of research and to encourage discussion.  The views expressed on statistical issues are those of the authors and not necessarily
those of the U.S. Census Bureau.}  }

\begin{document}
\maketitle

\newpage
\begin{abstract}
The paper provides a parametrization of Vector Autoregression (VAR) that enables one to look at the parameters associated with unit root dynamics and those associated with stable dynamics separately.
The task is achieved via a novel factorization of the VAR polynomial that partitions the polynomial spectrum into unit root and stable and zero roots via polynomial factors. The proposed factorization adds to the literature of spectral factorization of matrix polynomials.  The main benefit is that using the parameterization, actions could be taken to model the dynamics due to a particular class of roots, e.g. unit roots or zero roots, without changing the properties of the dynamics due to other roots. For example, using the parameterization one is able to estimate cointegrating space with appropriate rank that  maintains the root structure of the original VAR processes or one can estimate a reduced rank causal VAR process maintaining the constraints of causality. In essence, this parameterization provides the practitioner an option to perform estimation of VAR processes with constrained root structure (e.g., conintegrated VAR or reduced rank VAR) such that the estimated model maintains the assumed root structure. \\

\end{abstract}

\clearpage\pagebreak\newpage \pagenumbering{arabic}

\section{Introduction}
Since Sims (1980), Vector Autoregression (VAR) has taken a central role in the modeling and empirical analysis of macroeconomics series.  The existence of features such as persistence (Friedman and Kuttner, 1992),   co-movement (Stock and Watson (1989) and Christiano and Ljungqvist (1988)),  and reduced dimensional dynamics  (Stock and Watson (2002, 2016))    have made the analysis of such series exciting,  and has generated  a vast literature on unit root, co-integrated and reduced rank VAR systems. 
  
The exact nature of VAR dynamics is determined by the spectral configuration (number and type of roots) of the VAR polynomial.  The  main types of roots that show up in applications are: (1) stable roots, i.e. roots with magnitude less than one in absolute value;  (2) unit roots; and (3) zero roots. Stable roots pertain to  causality of sub-processes;  unit roots lead to nonstationary/integrated processes; zero roots  indicate reduced rank processes.  In this article we focus on  two main applications of these root structures:
  co-integrated VAR and reduced rank VAR.  In co-integrated VAR the root structure is a mixture of stable roots and roots equal to unity.  

A  common practice in co-integrated processes  is to study dynamics driven by the  stable roots  after differencing the data to remove the unit roots.    However, there have
  been debates about the appropriate way of handling the integrated part of the process and the consequences of over and under differencing;  see Toda and Yamamoto (1995), Ashley and Verbrugge (2009). There are existing models,  such as Vector Error Correction Models (VECM), that lead to efficient estimation of the cointegrating relations. Similarly, for reduced rank VAR there are methods, such as constrained least squares (L\"utkepohl,  2007) that can estimate the reduced rank structure.

Properly parametrizing the stable portion of a VAR polynomial is also important in Bayesian VAR modeling, a tool that has become popular for analyzing large systems of macroeconomic series; see  Koop (2013),  Chan et al. (2016), Carriero et al. (2016), and Koop (2017).  However, there is currently no scheme  for specifying a prior that is fully supported on the constrained parameter space described by a  co-integrated VAR --  see Doan et al. (1984), Ba\'nbura et al. (2010), and Giannone et al. (2015) for prior work. Recent literature on Bayesian co-integration has focused on estimation of the    co-integration space, and there the differenced form  (or the VECM form) 
  is conveniently    used to specify priors on the co-integrating space, identified with the Stiefel manifold. One of the motivational goals of this article is  to provide  a constrained  prior specification for co-integrated VAR.  

Differencing the data first and then  estimating stable relationships can be thought of as a sequential factorization  of the VAR polynomial, where one of the factor is the differencing operator $I - B$ ($B$ is the backshift operator and $I$ is the identity matrix),   and another factor is the stable part which involves the co-integrating relations.   Whether estimated as a VECM or a two stage model, current methods do not guarantee that the estimated full VAR model is constrained to have only unit roots and stable roots. This  introduces non-invertibity, an undesirable feature, for some of the   linear combinations of the differenced process.

Another application where the VAR root structure is constrained  is that of reduced rank VAR (Reinsel (1983), Velu, Reinsel and Wichern (1986),  and  Reinsel and Velu (1998)).  A reduced rank formulation  also  provides a parameterization for sparse VAR (Davis et al. (2015) and Koop and Korobilis (2015)).
Here the roots of the VAR polynomial  are a mixture of zero and nonzero stable roots, where the nonzero roots   correspond to latent casual autoregressive processes. The steps of reduced rank factorization of the VAR coefficients  followed by estimation of parameters can be thought of as  separation of the zero roots from the nonzero stable roots, in terms of   VAR polynomial factors.  In general, there are estimation methods that would constrain the estimated VAR polynomial to have   the required number of zero roots (L\"utkepohl,  2007), but such methods do not constrain the remaining roots to be stable.   

Thus, current methods for estimation of VAR models  do not restrict the estimates of the VAR coefficients to the constrained parameter space prescribed by the models. A suitable parameterization of the constrained VAR models can facilitate constrained parameter estimation. A parameterizaton would mean a one-to-one mapping, from the constrained model parameter space to an unrestricted Euclidean space, which  can be used to perform estimation of model parameters without the complication of parameter restrictions. The main objectives of this paper are:
\begin{enumerate}
\item
Provide  identifiable factorizations for VAR polynomials where the roots associated with polynomials factors are either stable or zero or one. 
\item
Use the factorization to provide an  identifiable parameterization of VAR processes for co-integrated and reduced rank VAR models. 
\end{enumerate}

\section{Co-integrated   VAR(1) Processes}

\subsection{Factorization}

To fix ideas we begin with the first order VAR process. 
Consider the $m$-dimensional VAR(1) process defined  by $\Phi(B) Y_t = Z_t$,  where $\Phi(B) = I_m - \Phi B$ is such 
 that the  eigenvalues of $\Phi$ are either unity or less than one in absolute value, and $I_m$ is the $m$-dimensional identity matrix.
 We will assume that $\mbox{rank} (\Phi) = m$ with $0 < r < m$  unit roots and  rank of $\Phi - I_m $ is $m-r$.
  (This implies that the unit eigenvalue is regular, i.e., the algebraic and geometric multiplicities of the unit root are the same.)
Define an $m$-dimensional {\it difference operator} of  rank $r \leq m$ via $\Delta_U(B) = ( I - UB)$, 
  where $U$ is a rank $r$ idempotent matrix and $B$ is the backshift operator. 
For an $m\times m$ matrix $\Phi$ having $r$ regular unit roots and $m - r$ stable roots,  define the class of left  factorizations 
\begin{equation}
 \mathcal{C}^L_{\Phi} = \{ (\Upsilon,U):  \Phi(B) = (I - \Upsilon B)\Delta_U(B)\}, 
\label{eq:left-factorization}
\end{equation}
such that the roots of the determinantal equation $\det(\Upsilon(B)) = 0,$ based on the first factor $\Upsilon(B) = I_m - \Upsilon B,$  are  either zero or are 
  the stable roots of  $\Phi$, and  $\Delta_U(B)$ is a  difference operator of rank $r.$ The zero eigenvalue of $\Upsilon$ is regular and  $\mbox{rank} (\Upsilon) = m - r.$ 
Necessarily, $\Upsilon \, U = 0$ and $\Upsilon + U = \Phi$.  Since the unit root 
 eigenvalue is assumed to be regular, $U$ is  necessarily idempotent, i.e., $U^2 = U$.

 Unlike the factorization of the univariate AR polynomials in terms of unit and stable roots, the matrix version is more subtle due 
  to non-commutativity of arbitrary matrices. Thus, we define the   corresponding right factorization class  via
\begin{equation}
 \mathcal{C}^R_{\Phi} = \{(\Upsilon ,U):  \Phi(B) = \Delta_U(B)(I - \Upsilon B)\}. 
\label{eq:right-factorization}
\end{equation}
Both the left and the right factorizations  $\Phi \leftrightarrow (\Upsilon, U)$ are closed
 under the group of similarity  transformations $(\Upsilon, U) \leftrightarrow (Q\Upsilon Q^{-1}, QUQ^{-1})$,  where $Q$ is a nonsingular matrix
 such that $Q \,  \Phi = \Phi \, Q.$

Factorizations such as \eqref{eq:left-factorization} or \eqref{eq:right-factorization}  may be employed in the context of co-integration. 
Suppose $\Pi := \Phi - I_m = \alpha \, \beta^{\prime}$  where $\alpha, \beta$ are $m \times (m-r)$ matrices of full column rank, 
  hence providing a rank factorization of $\Pi. $ 
The original VAR(1) written in the traditional vector error correction model (VECM) form is 
\begin{equation}
 (I_m - B) \, Y_t = \Pi \, B  \, Y_t + Z_t = \alpha \, \beta^{\prime} \, Y_{t-1} + Z_t,
\label{eq:VAR1_VECM}
\end{equation}
 where $\beta$ contains the cointegrating relations.

A  problem with taking the full difference $I_m - B$ is that the resulting process has noninvertible subprocesses.  
   While this does not create a direct problem in identification of the co-integrating relations, 
 noninvertibility may be problematic in other operations such as signal extraction (McElroy and Trimbur, 2015), 
 forecasting  (McElroy and McCracken, 2017), and estimation of structural shocks. 
 Instead, one could work with the reduced difference $\Delta_U(B) \, Y_t$, resulting in an invertible 
process while maintaining the same co-integrating relationship and the same error process. 
Specifically, if $X_t = \Delta_U(B) \, Y_t$ then $X_t$ satisfies the reduced rank VAR(1) given by
 $ X_t =  \Upsilon \, X_{t-1} + Z_t$  (if $ (\Upsilon,U) \in   \mathcal{C}^L_{\Phi}$) such that
 $\beta^{\prime} \, X_t$ is a stable VAR(1), i.e.,  
\[ 
 \beta^{\prime} \,  X_t = A \, \beta^{\prime} \, X_{t-1} + \beta^{\prime} \, Z_t
\]
with $A = \beta^{\prime} \, \alpha$ having all stable roots. 

In general, the class of left factorizations  and the class of right factorizations are different with empty intersection, but there exist bijections (see below) 
   from the left class to the right class, and hence parameterization of one class will automatically induce parameterization of the other class.    In the higher order case,  processes that admit factors that commute -- and hence can be considered both a left and a right factorization -- belong to a very restricted class of processes, and are of no  practical use.     However, in the first order VAR, one can   parameterize the entire co-integration space by demanding that    the stable fator $\Upsilon$ and the difference operator $\Delta_U$ commute and uniquely identify the parameters $\Upsilon, U$ using this requirement.  The following result shows that there is a unique factorization such that $I_m - \Upsilon B$ and $\Delta_U(B)$ commute,
   and the pair is invariant under  similarity transformations of the form 
\[
 (\Upsilon, U) \leftrightarrow (Q \, {\tilde{\Phi}} \, Q^{-1}, Q \, U \, Q^{-1})
\] 
 for any  nonsingular $Q$ that commutes with $\Phi$. The result is important because it facillitates the identification of the pair $(\Upsilon, U).$ 

\begin{Proposition}
\label{prop:commuting-decomp}
For the class of factorizations \eqref{eq:left-factorization} and \eqref{eq:right-factorization}, there exists  a unique pair    $({\bar{\Upsilon}},\bar{U}) \in \mathcal{C}_{\Phi}^L \bigcap \mathcal{C}_{\Phi}^R$,  and for  that pair $({\bar{\Upsilon}},\bar{U}) \leftrightarrow (Q \, {\bar{\Upsilon}} \, Q^{-1}, Q \, {\bar{U}} \, Q^{-1})$   for every nonsingular matrix $Q$ such that $Q \, \Phi = \Phi \, Q.$
\end{Proposition}
This is a desirable result, since the operation of differencing the process and fitting a stable VAR to the   co-integrating relations can be done interchangeably.

One could generalize the representation to other  unit root processes, such as a seasonal unit root. Suppose $\Phi$ has $r$ pairs of    complex conjugate roots equal to $e^{\pm i\theta}$  and assume that the roots are regular in the sense that   the geometric multiplicity is same as the algebraic multiplicity. Let
\[ 
 S(\theta) = \begin{bmatrix} \cos(\theta) & -\sin(\theta) \\ \sin(\theta) & \cos(\theta) \end{bmatrix}
\]
 be the orthogonal rotation matrix corresponding to $\theta.$ Then to parameterize processes 
  with monthly seasonal unit roots at $\theta = 2\pi j/12, \; j = 1,2,\ldots, 5$ (for example), one could use the factorization 
\[ 
 I_m - z \, \Phi = (I_m  - z \, \Upsilon) \, (I_m - z \, U), 
\]
where 
\[
 \Upsilon = P \, \begin{bmatrix} 0 & 0 \\ 0 & \Lambda \end{bmatrix} \, P^{-1}  \qquad 
U = P \, \begin{bmatrix} I_r \otimes S(\theta) & 0 \\ 0 & 0 \end{bmatrix} \, P^{-1},
\]
 with $\Lambda$ associated with the Jordan block for the stable roots of $\Phi$.  Note that in this case $U$ is a periodic matrix with $U^{12} = U$.  Thus, instead of idempotent difference matrices, for seasonal roots one may consider periodic matrices of appropriate seasonal period.   However, characterization and parameterization of such matrices needs to be investigated.

\subsection{Parameterization}
The parameterization of  co-integrated VAR(1) processes can be achieved by parameterizing the pair $(\Upsilon, U)$ and the error variance matrix $\Sigma$.  Since $U$ can be uniquely derived from $\Upsilon$, we must    parameterize the reduced rank matrix $\Upsilon$, while imposing  the constraint that the spectral radius is less than one.  We build upon the parameterization proposed in Roy et al. (2019), which     showed that an $m\times m$ non-singular matrix $\Upsilon$  is Schur-Stable (has only stable roots)  if and only if it can be represented as 
\begin{equation}
\Upsilon = V^{1/2} \, Q \, (I_m + V)^{-1/2},
\label{eq:VAR(1)_fullrank}
\end{equation}
where $V$ is an arbitrary positive definite matrix, $Q$ is an orthogonal matrix, and $A^{1/2}$ denotes the matrix square 
 root of $A$. The representation was obtained from the fundamental Riccati equations that the matrices must satisfy, namely 
\begin{equation}
\label{eq:riccati}
\Gamma =\Upsilon \, \Gamma \, \Upsilon^{\prime} + I_m,
\end{equation}
 where $\Gamma \geq I_m$ is a positive definite matrix and $V$ in the representation is $V = \Gamma - I_m$. 
 In other words, the matrices must satisfy $I_m + V = \Upsilon \, (I_m + V) \, \Upsilon^{\prime} + I_m, $ which leads   to the equation $V =  \Upsilon \, (I_m + V) \, \Upsilon^{\prime}$ and hence $V^{1/2} \, Q = \Upsilon \, (I_m + V)^{1/2}$.   In the present context, $\Upsilon$ is reduced rank or rank $r \leq m$.  In view of the representation \eqref{eq:VAR(1)_fullrank},  $\Upsilon$ is reduced rank if  and only if $V$ is reduced rank. To get a parameterization that is full rank, we need to reduce the dimension of $V$,    and hence that of $Q$. The following result provides such a full rank parameterization. 
\begin{Proposition}
\label{prop:schur-redrank-var1}
An $m\times m$ matrix $\Upsilon$ is Schur-stable and of rank $r$ if and only if there exists $m\times m$ nonnegative definite matrix $V$ of rank $r$ and $r\times m$ semi-orthogonal matrix $Q$  (i.e., $Q \, Q^{\prime} = I_r$), such that
\begin{equation}
\Upsilon = V_1 \, Q  \, (I_m + V_1 \, V_1^{\prime})^{-1/2},
\label{eq:VAR(1)_reducedrank}
\end{equation}
where   $V_1$ is an $m\times r$  square root of $V$,  and $(I_m + V)^{-1/2}$ is a square root of ${(I_m + V)}^{-1}.$ 
\end{Proposition}

There are   $mr - r(r-1)/2$ free parameters in $V_1$ and $mr - r(r+1)/2$  free parameters in $Q$. Thus the total number of free parameters in the representation is $d_r = r(2m - r).$ Note that, in the full rank case $d_m = m^2$.  This is obviously known for general reduced rank matrices.   Specifically, $\Upsilon$ belongs to the rank manifold
\begin{equation}
\mathcal{M}_r = \{ M \in \mathbb{R}^{m\times m}: \mbox{rank} (M) = r \}.
\label{eq:rank_manifold}
\end{equation}
The manifold has co-dimension $(m - r)^2$ (Guillemin and Pollack, 2010),  and hence the dimension is  $d_r = m^2 - (m - r)^2 = r(2m - r).$ Thus, the reduction from the reduced rank constraints is $(m - r)^2$,    which can be significant depending on $m$ and $r$.

\section{The Co-integrated VAR($p$)}

\subsection{Factorization}

We discuss parameterization of the  $p$th order co-integrated VAR process. Parameterization of the reduced rank process
  is obtained as an intermediate step of the full co-integrated parameterization.
Since the analysis of the VAR($p$) case revolves around the properties of the VAR($p$) polynomial, we first develop  notation for classes   of matrix polynomials.  For a  matrix  polynomial $A(z) =  I_m - A_1 z - \cdots - A_k z^k, $ we will denote  by ${\tilde{A}}(z) = z^kA(z^{-1}) = z^k I - A_1 Z^{k-1} - \cdots A_{k-1} z  - A_k$ the corresponding monic matrix polynomial.  A matrix  polynomial $A(z) =  I_m - A_1 z - \cdots - A_k z^k, $ will be called {\it Schur-stable} if all roots 
 of $\det {\tilde{A}}(z) = 0$ lie in the interior of  the  unit disc $\mathcal{D} = \{z \in \mathbb{C}: |z| < 1\}.$  Let
\[
  \mathfrak{S}^{m,k} =  \{  A(z) = I_m - A_1 z - \cdots - A_k z^k: A_j \in \mathbb{R}^{m\times m},  
   A_k \ne 0 \mbox { and } A(z) \mbox{ is Schur-stable}\} 
\]
define the set of all $m-$dimensional Schur-stable matrix  polynomials of degree $k$ with the constant term as the identity matrix.  Along with the Schur-stable  $m$-dimensional polynomials, we define the extended Schur-stable  polynomials $A(z) = I_m - A_1  z - \cdots - A_k z^k$  as polynomials where the roots of $\det {\tilde{A}}(z) = 0$ are either unity or are inside the  unit disc.  A sub-class of such polynomials, where some of the roots   of $\det {\tilde{A}}(z)  = 0$  are exactly zero,  will be called {\it reduced rank} polynomials.  Specifically, define the class of  {\it extended Schur-stable}  polynomials via
\begin{eqnarray}
{\bar{\mathfrak{S}}}^{m,k}_{r,s} &=& \{  A(z) = I_m - A_1 z - \cdots - A_k z^k: A_k \ne 0,\; det {\tilde{A}}(z) = 0 \mbox{ has }  r \mbox{ roots equal to 1}  \nonumber \\
&&  \qquad \qquad\qquad \qquad   \mbox{ and } s \mbox{ roots equal to zero, and the rest within } \mathcal{D} \}.  
\end{eqnarray}
Also denote ${\bar{\mathfrak{S}}}^{m,k}_r = \bigcup_s {\bar{\mathfrak{S}}}^{m,k}_{r, s}, $   ${{\mathfrak{S}}}^{m,k}_s = \bigcup_r {\bar{\mathfrak{S}}}^{m,k}_{r, s}$ and ${\bar{\mathfrak{S}}}^{m,k} = \bigcup_{r,s} {\bar{\mathfrak{S}}}^{m,k}_{r, s}.$
Note that ${\bar{\mathfrak{S}}}^{m,k}_0 = {{\mathfrak{S}}}^{m,k}.$ 

The main goal   is factorization of the VAR($p$) polynomial into suitable factors  such that the unit roots and the stable root separate. Factorization of matrix polynomials,  with the roots of the factors forming a partition of the spectrum of the polynomial,  has been extensively studied in the literature. The theory of  such spectral factorization  generally depends  on the properties of the Jordan triplet associated with the polynomial (see Goldberg et al. (1982)).  We want to separate the unit root part of the spectrum from the stable roots  in such a way that this separation is  expressed   in terms of polynomial factors, e.g. a difference operator and a stable operator.  However, there is one key distinction that makes the existing theory inapplicable  to the present situation. The factorization of the spectrum is generally done in multiples  of the dimension, e.g., if there is an $m$-dimensional matrix polynomial with $mk$ roots, then 
 in the classical treatment  the number of roots in partition of the spectrum is a multiple of $m$.   In our case, the number of unit roots need not be a multiple of $m$.    Hence, as in the first order case, we need to augment  the factors with zero roots   to make the number of roots a multiple of $m$. Augmentation by
  zero roots does not   change the intrinsic nature of the process. 
 
The number of unit roots, $r$, for a $p$th degree $m$-dimensional polynomial could be as high as $mp$,   but for most interesting applications of co-integrated processes where the first difference is a stationary process,   we restrict to the case $0 \leq r < m.$   For a polynomial $A(z) =  I_m - A_1 z - \cdots - A_k z^k, $   let $C_A$ denote the {\it companion } matrix of ${{A}}$; it is known that Schur-stability of $A(z)$ is equivalent to  the eigenvalues of $C_A$ lying within $\mathcal{D}$ (cf. L\"utkepohl, 2007). A root of $\det {\tilde{A}}(z)  = 0$ will be called {\it regular} (Banerjee and Roy, 2014) if its algebraic multiplicity   is same as its geometric multiplicity, as an eigenvalue of $C_A;$   Throughout the paper we will assume that both the   unit roots and  the zero roots of    any polynomial in ${\bar{\mathfrak{S}}}^{m,k}_{r,s}$ are all regular.  The number of zero roots for a polynomial in ${\bar{\mathfrak{S}}}^{m,k}_{r,s} $ is restricted to $0 \leq s < m.$  For $s > 0$ we have the following straight-forward result.

\begin{Proposition}
\label{prop:reduced_last_coeff}
Let $A(z) =  I_m - A_1 z - \cdots - A_k z^k.$  The number of zero roots of $\det {\tilde{A}}(z) = 0$, say $r,$ is greater   than zero  if and only  if $\det A_k = 0$, and hence $A_k$ is reduced rank. 
\end{Proposition} 

Thus, for our setup,  the class of reduced rank polynomials is also the class of polynomials where the constant matrix is reduced rank.  This fact could be used in parameterization of the class of reduced rank polynomials.   We  propose a   factorization of  VAR($p$) polynomials $\Phi (z) \in {\bar{\mathfrak{S}}}^{m,k}_{r,0}$ into a purely causal  VAR($p$)  factor $\Upsilon(z) \in {{\mathfrak{S}}}^{m,k}_r$ and a difference operator $ \Delta_U(z)$,  yielding the decomposition
\begin{equation}
   \Phi (z) = \Upsilon (z) \, \Delta_U(z).
\label{eq:leftfact}
\end{equation}
Here, the monic versions of $\Upsilon (z)$ and $\Phi (z)$, namely $\tilde{\Upsilon} (z)$ and $\tilde{\Phi} (z)$, have the   property that  the $(m-r)$ nonzero roots of  ${\tilde{\Upsilon}} (z)$  are the same as the $(m - r)$ stable  roots ${\tilde{\Phi}} (z)$,  with the same exact multiplicities; also,  $U$ is a diagonalizable matrix with $r$ unit roots  and $(m-r)$ zero roots, and hence can serve as a difference operator.  The following theorem summarizes these results. 
\begin{Theorem}
Given  $\Phi(z) \in {\bar{\mathfrak{S}}}^{m,k}_{r,0}$  there  exists unique   $\Upsilon(z) \in {\mathfrak{S}}^{m,k}_r$  and a   rank $r$ symmetric idempotent matrix $U$  such that (\ref{eq:leftfact}) holds,  and  the nonzero stable roots of $\det {\tilde{\Upsilon}}(z) = 0$ are the $(m - r)$ stable roots of $\det {\tilde{\Phi}}(z) = 0.$  Conversely, given  $\Upsilon(z) \in {\mathfrak{S}}^{m,k}_r$, there exists unique $\Phi(z) \in {\bar{\mathfrak{S}}}^{m,k}_{r,0}$ and a  rank $r$ symmetric idempotent matrix $U$    such that  (\ref{eq:leftfact}) holds,  and the  stable roots of $\det {\tilde{\Phi}}(z) = 0$ are the $(m - r)$ nonzero roots of $\det {\tilde{\Upsilon}}(z) = 0.$
\label{thm:varp}
\end{Theorem}

One could similarly state a theorem for a right factorization of the form 
\begin{equation}
\Phi(z) = \Delta_U(z) \, \Upsilon(z),
\label{eq:rightfact}
\end{equation}
where $U = P_{\mathcal{N}(\Pi^{\prime})} = P_{\mathcal{N}(\Upsilon_k^{\prime})}.$ 
Such a factorization is also unique provided one demands $U$ is a symmetric idempotent  matrix of appropriate rank. However, one could state a more general result regarding the class or left and right factorization.  Specifically, if one only restricts to idempotent difference operator $U$, then there is a bijection between the space   of left factorizations and the space of right factorizations. Note that once $\Phi$ is given, then the stable factor   $\Upsilon_L(z)$ can be solved in terms of the difference matrix $U$ and $\Phi$  using the relation  (\ref{eq:Upsilon_from_phi})    for the left factorization,  and the stable factor $\Upsilon_R(z)$ has an analogous relationship for the right factorization.   Thus, given $\Phi$ one could denote the factors as $(\Upsilon, U) \doteq (\Upsilon(\Phi, U), U).$ 
\begin{Theorem}
\label{thm:factor-leftright}
Suppose  $\Phi(z) \in {\bar{\mathfrak{S}}}^{m,k}_{r,0}$ is given. Let the classes, $\mathcal{F}_L(\Phi) $, $\mathcal{F}_R(\Phi),$ defined as 
\begin{eqnarray}
\mathcal{F}_L(\Phi) &=& \{ (\Upsilon(z), U):  \Phi(z) = \Upsilon(z)\Delta_U(z),\; \Upsilon(z) \in {\mathfrak{S}}^{m,k}_r, U^2 = U,  \mbox{rank} (U) = r \} \nonumber \\
\mathcal{F}_R(\Phi) &=& \{ (\Upsilon(z), U):  \Phi(z) = \Delta_U(z)\Upsilon(z),\; \Upsilon(z) \in {\mathfrak{S}}^{m,k}_r, U^2 = U, \mbox{rank} (U) = r \},
\label{eq:factor_class}
\end{eqnarray}
be the class of left and right factorizations of $\Phi(z),$ respectively. Then there is bijection $T_{\Phi}$ from 
\[
 \mathcal{F}_L(\Phi)  \stackrel{T_{\Phi}}{\longrightarrow} \mathcal{F}_R(\Phi) 
\]
given by $ (\Upsilon_L(\Phi, U), U) \to (\Upsilon_R(\Phi, T_{\Phi}(U)), T_{\Phi}(U))$, where $T_{\Phi}$ is a linear map. 
\end{Theorem}

\subsection{Parameterization}
Theorem~\ref{thm:varp} shows that the required parameterization of the polynomials in  co-integrated models can be done using reduced rank polynomials $\Upsilon.$   Along with Proposition~\ref{prop:reduced_last_coeff} this indicates the need for polynomials with reduced rank constant term.  In this sub-section we first prove a fundamental result on the representation of reduced rank   VAR($p$) polynomials with stable roots. The result will help us describe the required parameterization.  We first introduce notation from Roy et al. (2019) that established a parameterization for the VAR($p$) Schur-stable space.   For $j \geq 1,$ define ${\underline{U}}_j$ to be a symmetric block Toeplitz matrix of order $j$:
\begin{equation}
{\underline{U}}_j = \begin{pmatrix}
U(0) & U(1) & \cdots & U(j) \\
U(1)^{\prime} & U(0) & \cdots & \cdot              \\
\vdots        &  \ddots        & \ddots & \vdots             \\
U(j)^{\prime} & \cdots     & U(1)^{\prime} & U(0)
\end{pmatrix}
\label{blocktoeplitz}
\end{equation}
where $U(0), U(1), \ldots, U(j)$ are arbitrary $m\times m$ matrices and $U(0) \in {\mathscr{S}}^{m} $. Note that ${\underline{U}}_0 = U(0).$  We will take advantage of the following nested representations of ${\underline{U}}_j$ in terms of ${\underline{U}}_{j-1}$: the lower representation given by 
\begin{equation}
{\underline{U}}_j = \begin{pmatrix}
U(0) & \xi_j^{\prime} \\
\xi_j & {\underline{U}}_{j-1}
\end{pmatrix},
\label{lower_ut}
\end{equation}
and the upper representation given by 
\begin{equation}
{\underline{U}}_j =  \begin{pmatrix}
{\underline{U}}_{j-1} & \kappa_j \\
\kappa_j^{\prime} & U(0)
\end{pmatrix}.
\label{upper_ut}
\end{equation}
Here $\xi_j^{\prime}  = (U(1), \cdots, U(j))$ and $\kappa_j^{\prime} = ({U(j)}^{\prime},\ldots, {U(1)}^{\prime} ).$  
 The Schur complements of ${\underline{U}}_{j-1}$ in ${\underline{U}}_{j}$ in the two representations \eqref{lower_ut} and \eqref{upper_ut} are
\begin{equation}
C_j = U(0) - \xi_j^{\prime}{\underline{U}}_{j-1}^{-1}\xi_j,
\label{lower_schur}
\end{equation}
\begin{equation}
D_j = U(0) - \kappa_j^{\prime}{\underline{U}}_{j-1}^{-1}\kappa_j.
\label{upper_schur}
\end{equation}
Define $C_0 = D_0 = U(0).$ Let $\mathfrak{T}^{m,k}$ denote the set of $m(k+1)\times m(k+1)$ symmetric block Toeplitz matrices with $m$-dimensional blocks, i.e.,
\[ 
{\mathfrak{T}}^{m,k} = \{ {\underline{U}}_k \in  {\mathscr{S}}^{m(k+1)} : {\underline{U}}_k  \mbox{ is in the form } \eqref{blocktoeplitz} \}. 
\]
Also define $ {\mathfrak{T}}^{m,k}_{++}$  to be the subset of $ {\mathfrak{T}}^{m,k}$ comprising the positive definite block Toeplitz matrices of   order $k$ and $m$-dimensional blocks.  Then Roy et al. (2019) shows that every Schur-stable $k$th order $m$-dimensional matrix polynomial   $z^k I_m - \sum_{j=1}^k z^{k-j} \Upsilon_j$ can be written as $[\Upsilon_1^{\prime}, \ldots, \Upsilon_k^{\prime}] = \xi_{k}^{\prime} \, {\underline{U}}_{k-1}^{-1}$;    also, a symmetric block Toeplitz matrix $U$ is positive definite  if and only if  $C_0 \loeq  C_1 \loeq \cdots \loeq C_k \loe 0$, where $\loe$ indicates the Lowner ordering. The parameterization for the Schur-stable space  was accomplished noting that the successive Schur complement differences $C_{j-1} - C_j$ are unrestricted positive semi-definite matrices.

The number of zero  eigenvalues (algebraic multiplicity) of $\Upsilon_k$ is less than or  equal to $r$, the number of zero   roots of $\det \Upsilon(z) = 0.$  For parameterization we will assume $r \leq m$.
 Otherwise, barring some uninteresting cases (e.g., $\Upsilon_k$ is a nilpotent matrix)   the processes can be represented as a lower order process. Given that  $ 0 \leq r \leq m,$  we will focus on $k$th degree $m$-dimensional polynomials in $\mathfrak{S}^{m,k}_r$ with  $\mbox{rank} (\Upsilon_k) = (m - r)$.   A measure theoretic argument can be made to show that this class is  within  $\mathfrak{S}^{m,k}_r$.
%
Thus, without loss of generality,  we will parameterize  Schur-stable reduced rank polynomials where the rank reduction is entirely due to the constant matrix $\Upsilon_k.$  The following theorem provides a representation of such reduced rank Schur-stable polynomials. 

\begin{Theorem}
  \label{thm:schur_iter}
A block Toeplitz matrix ${\underline{U}}_k \in \mathfrak{T}^{m,k}$ is positive definite, and the associated Schur-stable polynomial  $\Upsilon(z) = z^k I_m - \sum_{j=1}^k z^{k-j} \Upsilon_j$ is reduced rank in the sense that $\mbox{rank} (\Upsilon_k) = (m-r)$,    if and only if the associated Schur complement sequence $C_j = U(0) - \xi_j^{\prime}{\underline{U}}_{j-1}^{-1}\xi_j$  satisfies $C_0 \loeq  C_1 \loeq \cdots \loeq C_k \loe 0$ with $rank(C_{k-1} - C_k) = m - r$. 
\end{Theorem}

To complete the parameterization, following Roy et al. (2019) we will write $V_j = C_{j-1} - C_j$ and iteratively solve for    $U(j)$ from those differences. Following Theorem~ \ref{thm:schur_iter}, $V_1, \ldots, V_{k-1}$ are positive definite   and $V_k$ is positive semi-definite with rank equal to that of $\Upsilon_k$, where  $ V_j = B_j \, B_j^{\prime}$ and  
\[ 
B_k =  (U(k) - \xi_{k-1}^{\prime}{\underline{U}}_{k-2}^{-1}\kappa_{k-1}) \, D_{k-1}^{-1/2}.
\]
The following result is needed to establish the required relation between the Schur complement difference $V_j$  and $U(j)$.
\begin{Proposition}
\label{prop:ortho-matrix}
Let $A$ be an $m\times p$ matrix of rank $p$, and let $B$ be an $m\times m$ matrix of rank $p$.   Then $A \, A^{\prime} = B \, B^{\prime}$ if and only if   there exists an $p\times m$  semi-orthogonal matrix $Q$ with $Q \, Q^{\prime} = I_p,$ such that $A \, Q = B.$ 
\end{Proposition} 

As a result
\[
 V_j^{1/2} \, Q_j = B_j = (U(j) -  \xi_{k-1}^{\prime}{\underline{U}}_{k-2}^{-1}\kappa_{k-1}) \, D_{k-1}^{-1/2},
\]
where $V_j = V_j^{1/2} \, {V_j^{1/2}}^{\prime}$ and $Q_j$ is orthogonal or semi-orthogonal depending   on whether $V_j$ is full rank or reduced rank, respectively. Then 
\[
 U(j) =  \xi_{k-1}^{\prime}{\underline{U}}_{k-2}^{-1}\kappa_{k-1} + V_j^{1/2} \, Q_j \, D_j^{1/2}.
\]
Thus, one could iteratively reconstruct the $U(j)$ once the $V_j$s and $Q_j$s have been specified.  This provides the inverse map from the set of  positive definite matrices and orthogonal matrices to the set   of positive definite $m(k+1) \times m(k+1)$  block Toeplitx matrices ${\underline{U}}_k$,  with the last Schur    complement difference $C_{k-1} - C_k$ having rank  = $(m - r)$. Once the block Toeplitz matrix has been reconstructed,    the coefficient matrices can be obtained as the Yule-Walker solution corresponding to ${\underline{U}}_k$.

\subsection{ VAR$(p)$ computation and simulation}
One of the main goal of this paper to provide a parameterization that will alleviate the need to do likelihood based inference under complicated constraints on the parameter set. We can write down the likelihood for the transformed parameters in the co-integrated model using the proposed parameterization. The main idea is that since we can derive the difference operator $U$ in terms of the transformed parameters of $\Upsilon$, given a value of the transformed parameter we can difference the data using $U$ and write a stable reduced rank likelihood for $\upsilon$ in terms of the differenced data.  Suppose the observed process is 
\begin{equation}
X_t = \eta_t + Y_t,
\label{eq:constant_term}
\end{equation}
 where $\Phi(B)Y_t = Z_t$ is the mean zero VAR($k$) process with polynomial $\Phi(B) = \Upsilon(B)\Delta_U(B)$ and $\eta_t$ is a deterministic time varying mean process for $X_t$. Assume that $\Delta_U(B)\eta_t = c$ where $c$ is some constant  vector. Hence the derived model for $X_t$ is 
\[ \Phi(B)X_t = \mu + Z_t, \]
where the constant term in the model $\mu = \Upsilon(1)c$ and $\Upsilon(1) = I_m - \sum_{j=1}^k \Upsilon_j$ is a nonsingular matrix by assumption.   Thus, if $V_t = \Delta_U(B) X_t,$ then $V_t$ is a causal VAR($k$) process with a constant term $\mu.$ Note that by assumption $\eta_t = (I_m +(t-1)U) \, \Upsilon^{-1}(1) \, \mu$ gives a linear trend model for the time varying mean of the observed process.

We can write a reduced rank causal Gaussian  likelihood $\mathcal{L}_{RR}(\Upsilon(\eta) |  \mbox{data})$ with respect to parameters $\eta$
  that are associated with parameterization of $\Upsilon$ and data here is the differenced data $\Delta_{U(\eta)}(B) X_t.$ 

 In each simulation run we investigate  the integrated mean squared error (re-scaled by sample size),
   \[   
\mbox{MSE}_T  = T M^{-1} \, \sum_{k=1}^M \|\Upsilon^{(k)} - \Upsilon\|_F^2, 
\]
where $T$ is the sample size, $M$ is the number of Monte Carlo replications,  $\Upsilon^{(k)}$ is the estimator of the reduced rank coefficient matrix $\Upsilon$ for the $k$th Monte Carlo replication, and    $m^{-1}\|A - B\|_F$ is the root mean squared error metric with  $\| \cdot  \|_F$ denoting the Frobenius norm of an $m\times m$ matrix. 
  What we report is relative efficiency of the MLE with respect to the OLS estimators of $\Upsilon.$ We define the relative efficiency as  two estimators ${\widehat{\Upsilon}}_1$ and ${\widehat{\Upsilon}}_2$,    denoted by $ \mbox{Eff} ({\widehat{\Upsilon}}_1 | {\widehat{\Upsilon}}_2)$, as the ratio of  the $\mbox{MSE}_T $ of the two estimators.

To start the MLE iteration we also need an initial estimator that satisfies the stability and reduced rank constraints. There is no obvious choice.  There are several ways one could obtain such an initial estimator: (i) choose an unconstrained estimator  (such as the OLS estimator) of $\Upsilon$ and then modify the estimator to make it Schur-stable, and then project the solution to the reduced rank manifold  without changing the stability property; (ii) choose an unconstrained estimator,    project it to the reduced rank manifold, and then modify the solution to make it Schur-stable without destroying the reduced rank structure;    (iii) choose a stable estimator  (e.g., the  Yule-Walker estimator) and project it to the reduced rank manifold;     (iv) choose a reduced rank estimator such as reduced rank OLS and modify it to make it stable without changing the reduced rank structure.  We use option (i) for simulation.  \\

\noindent{\bf Initial estimation of transformed parameters}\\
 For initial parameters values for MLE iterations, we need estimates of $\mu$ and the transformed parameters corresponding to an initial estimate of $\Upsilon$. To do so one could use the following scheme: 
\begin{enumerate}
\item
Estimate $\mu$, $\Phi$ and $\Sigma$ using OLS estimator $\mu_{OLS}$, ${{\Phi}}_{OLS}$ and $\Sigma_{OLS}$ . From $\Phi_{OLS}$ we obtain the  associated ${{\Pi}}_{OLS}  =   \sum_{j-1}^k \Phi_{j,OLS} - I_m.$ 
\item
Construct an initial estimate of the factor $\Upsilon$,  say ${\widetilde{\Upsilon}}$ using $\Phi_{OLS}$ and the  relation (\ref{eq:Upsilon_from_phi}). However, this initial estimate  need not be stable and need  not have the appropriate number of zero roots.  To ensure that the estimator of $\Upsilon$ used  in the initial MLE iteration conforms to the assumption of stability and reduced rank,  we first stabilize  ${\widetilde{\Upsilon}}$  and then project the stabilized value to the relevant reduced rank manifold to obtain the initial estimator of $\Upsilon$. 
\item
To stabilize   ${\widetilde{\Upsilon}}$  we use the SHRINK algorithm  given in Roy et al. (2019) and obtain a stablized form ${\widetilde{\Upsilon}}_S$ for ${\widetilde{\Upsilon}}.$
\item
Next we project  the stabilized  estimator ${\widetilde{\Upsilon}}_S$ to the space of $m$-dimensional, $k$th degree stable polynomials with exactly $r$ zero roots for the associated determinantal equation using the following algorithm: 
\begin{itemize}
\item
For any positive definite $m\times m$ matrix $V$ with eigenvalues $\lambda_1 \geq \lambda_2 \geq \cdots \geq \lambda_m,$ and associated orthonormal eigenvectors $p_1, p_2, \ldots, p_m$ define the rank $(m-r)$ projection as $Proj_r(V) = \sum_{j=1}^{m-r} \lambda_j p_j p_j^{\prime}.$ 
Construct the associated $V$ and $Q$ parameters following the algorithm given in Roy et al. (2019). Modify the last pair $(V_k, Q_k)$ as: $V_k \to Proj_r(V_k)$ and $Q_k \to Q_{k,r}$ where $Q_{k,r}$ consists of the first $(m-r)$ columns of $Q_k$.

\item
Reconstruct the modified initial estimator $\Upsilon_{OLS}$ using the inverse algorithm  from $(V_1, Q_1), $ $  \ldots,  (V_{k-1}, Q_{k-1}), (Proj_r(V_k), Q_{k,r})$ and get the initial parameters associated with the modified $\Upsilon$ estimator using the algorithm defined in Roy et al. (2019). 
\end{itemize}
\end{enumerate}
The  MLE iteration can get stuck at the initial parameter values, and then     needs some perturbation to make it move over the transformed parameter space.      We then add small Gaussian noise (variance equal to 0.01) to every coordinate of the transformed parameter  vector; this is a restricted form of simulated annealing.  \\

\noindent{\bf Simulation}\\
We  investigate the utility of the proposed   parameterization in  co-integrated VAR  models.  For simulation we take the constant term $\mu_t$  to be known, and hence known to be zero without loss of generatlity. Consider the  following two- and  three-dimensional VAR($2$) processes:
\[ \Upsilon(B)(I - UB)Y_t  = Z_t,\]
where $Z_t$ are i.i.d. $N(0, \Sigma)$, $\Sigma = \tau^2 I_3,$
\begin{itemize}
\item
Case 1: $\Upsilon_1 = \begin{pmatrix} 0.2 & 0.1 \\ 0.1 & 0.2 \end{pmatrix}, \;\;\;\;\;\; \Upsilon_2 = 
\begin{pmatrix} 0.25 & 0.25 \\ 0.25 & 0.25 \end{pmatrix}.$
\item
Case 2:  $\Upsilon_1 = \begin{pmatrix} 0.2 & 0.1 & 0\\ 0.2 & 0.2 & -0.1\\ 0 & 0.05 & 0.1\end{pmatrix}, \;\;\;\;\;\; \Upsilon_2 = 
\begin{pmatrix} 0.1 & 0.1 & 0.1\\ 0.1 & 0.1 & 0.1\\ 0.1 &  0.1 & 0.1\end{pmatrix}.$
\item
Case 3:  $\Upsilon_1 = \begin{pmatrix} 0.8648 & 0.2313 & -0.4839\\ -0.8214 & 0.4676 & 0.5486\\ -0.9843 & -0.3214 & 1.0197\end{pmatrix}, \;\;\;\;\;\; \Upsilon_2 = 
\begin{pmatrix} -0.7093 & -0.0929 & 0.0610\\ 1.4071 & 0.4365 & -0.2552\\ -0.2234 &  0.5193 & -0.6245\end{pmatrix}.$
\end{itemize}
 In each case $U$ is the projection operator for the null space of $\Upsilon_2$. 

In Case~1, the polynomial ${\widetilde{\Upsilon}}(z) = z^2I_2 - z\Upsilon_1 - \Upsilon_2$ is reduced  rank with $\det \Upsilon(z) = 0$ having $r = 1$ zero root and $\mbox{rank} (\Upsilon_2)  = 1.$ 
 The four roots of $\det {\tilde{\Upsilon}}(z) = 0$ are $0.8728$, $0.5728$, $0.1$ and $0$, respectively.  In Case~2,  $\Upsilon(z) $ is reduced rank with $r = 2$ zero roots and  $\mbox{rank} (\Upsilon_2) =  1.$    In Case~3, we have a randomly generated VAR($2$) polynomial with two unit roots,   where rank of the corresponding $\Upsilon_2$ is $1$. The stable roots of the polynomial are two complex conjugate   roots of magnitude $0.9416$, and two real roots of magnitude $0.3525$ and $0.2011$.

 From Table~\ref{tab:sim_VAR2},  we can see that the efficiency  of the constrained maximum likelihood estimator of $\Phi$  is slightly better than that of the   OLS estimator of $\Phi$, while the MLE is substantially  better in terms of efficiency compared to Yule-Walker estimators, both for $\Phi$ and $\Upsilon$.  

\begin{table}[!htbp]
\centering
\caption{Relative efficiency of estimators in the co-integrated VAR models }
\begin{tabular}{lccc}
  \hline
Case & $\mbox{Eff} ({\widehat{\Upsilon}}| {\widehat{\Upsilon}}_{YW})$  & $\mbox{Eff} ({\widehat{\Phi}}| {\widehat{\Phi}}_{OLS}) $ & 
 $\mbox{Eff} ({\widehat{\Phi}}| {\widetilde{\Phi}}_{YW}) $  \\ 
\hline
Case 1: & 1.20 & 1.01 &  1.67  \\
Case 2: & 1.18 & 1.04 &  2.51  \\
Case 3: & 1.26 & 1.18 &  1.28\\
Case 4: & 1.55 & 1.42 &  2.21 \\
\hline
\end{tabular}
\label{tab:sim_VAR2} 
\end{table}

We also performed limited simulations on higher dimensional and higher order processes.   We   include the relative efficiency of the estimators   for a six-dimensional VAR($1$)  
 under Case~4 in Table~\ref{tab:sim_VAR2}.    We do not report the exact value of the VAR matrix to save space, but the exact matrix  used for simulation can be obtained from the authors. The number of unit roots of the VAR($1$)   matrix was chosen to be 4 and hence there were $20$ free parameters in the $\Phi$ matrix and 21 free parameters in the $\Sigma$ matrix,  rendering the  dimension of the parameter space equal to $41$. The two stable roots complex conjugate roots   $-0.3452 \pm i 0.7461.$ For higher dimensional processes the gains in efficiency even with respect to the unconstrained OLS estimator  can  be quite substantial. 

\section{Data analysis}

National income and product accounts (NIPAs) of the United States are published by the Bureau of Economic Analysis and provide quarterly estimates for the four
  major components of GDP: Personal expenditure, Exports, Gross Private Domestic Investment and Government spending. It is of interest to understand
 the joint temporal dynamic of these components of   GDP.  Different  components of the Gross Domestic Product show diffrent temporal dynamics,
 some being more volatile (e.g., Investment).    We analyze the quarterly seasonally adjusted Exports, Gross Private Domestic Investment, and Government Consumption Expenditures \& Gross Investment series  (Trillions of Dollar)  as a tri-variate VAR,  and look for a co-integrating relationship using the proposed parameterization. For analysis we choose a time span of 40 years (1999 Q2 -- 2018 Q1), and thus the sample size is $T = 160.$ 

\begin{figure}
\centering
\includegraphics[width = 3 in]{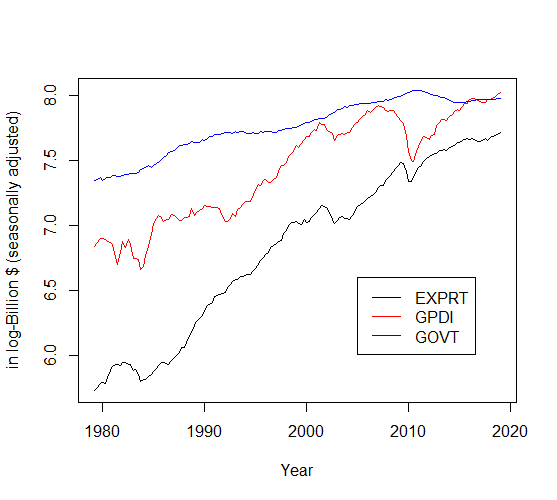}
\caption{Export, Gross Private Domestic investment and Goveernment expenditure (1999 Q2 - 2018 Q3) }
\label{fig:GDP}
\end{figure}
From the plot   (Figure~\ref{fig:GDP}), the three time series exhibit integrated behavior with possibly co-integrated relations. From univariate modeling for the series, it seems that a 
  lower order ARIMA($p$,1,0) fits well.  Thus, it seems natural to use a lower order VAR with possible unit roots. Indeed, after fitting and  residual analysis the VAR(4) model was the best candidate among contending VAR($p$) ($p \leq 6$)  models.  We present here the results from the VAR(4) modeling exercise for the GDP data.

To decide on the number of unit roots,  and hence the co-integrating rank of the series, we performed Johansen's test using the `CA.JO' function in the `URCA' package in R. The test statistic values for the  Johansen's  max-eigen test and Johansen's trace test were, respectively, (36.08,  17.14, 6.42) and (59.64, 23.56, 6.42). The first set of test values provide evidence  that the co-integrating rank is $H_0: r =  r^*$ versus the alternative $H_1: r = r^*+1$, for $r^* = 0,1,2,$  whereas the second set of values are associated with testing that the  co-integrating rank is $H_0: r = r^*$ versus it is $H_1: r^* < r \leq 3$  for $r^* = 0,1,2$. The critical values for the tests  at
 a nominal $5\%$     level  are (22.00,   15.67, 9.24) for the max-eigen tests and (34.91,19.96, 9.24)   for the trace test.   It is known that when the two tests do not agree in terms of the estimate of the co-integration rank, then the max-eigen test value is preferred (Dutta and Ahmed, 1997;  Odhiambo, 2005).  Thus, based on the observed values, the co-integrating rank was chosen to be two, i.e.,  the process is modeled with the restriction that the VAR polynomial has one unit root.  We used a constant term  in the mdoel. All estimates  were 
 rounded to three  decimal places.

The OLS estimate for the mean was $\mu_{OLS} = (0.010, 0.009, 0.004)^{\prime}.$  The MLE for the constant term in the model was $\mu_{MLE} = (-0.227, 0.187, 0.097)^{\prime}.$  The estimated polynomials using the OLS and MLE  with the given parametrization and the estimated error covariance matrices are given in the supplement. 
The absolute value of roots for the VAR polynomial for the OLS and MLE procedures were 
\begin{eqnarray*}
\lambda_{OLS} &=& (0.993, 0.945 ,0.945 ,0.799 ,0.799 ,0.603, 0.603, 0.555 ,0.555 ,0.413 ,0.413, 0.261), \\
\lambda_{MLE} &=& (1.000, 0.942, 0.942, 0.795, 0.795, 0.608, 0.608, 0.532, 0.532, 0.437, 0.437, 0.294).
\label{eq:estimated_roots_DOT}
\end{eqnarray*}

The OLS estimator is not restricted to have a unit root, and here estimates a causal process that is nearly nonstationary with two roots close to unity. The MLE with the  constraint that co-integration rank is two, estimates a root that is exactly equal to one and the remaining roots less than one in magnitude. Other than the  unit root,  MLE estimates one root with magnitude close to one. Thus, one of the two estimated co-integrated processes  is expected to be nearly nonstationary. 

\begin{figure}
\centering
\includegraphics[width = 2.5 in]{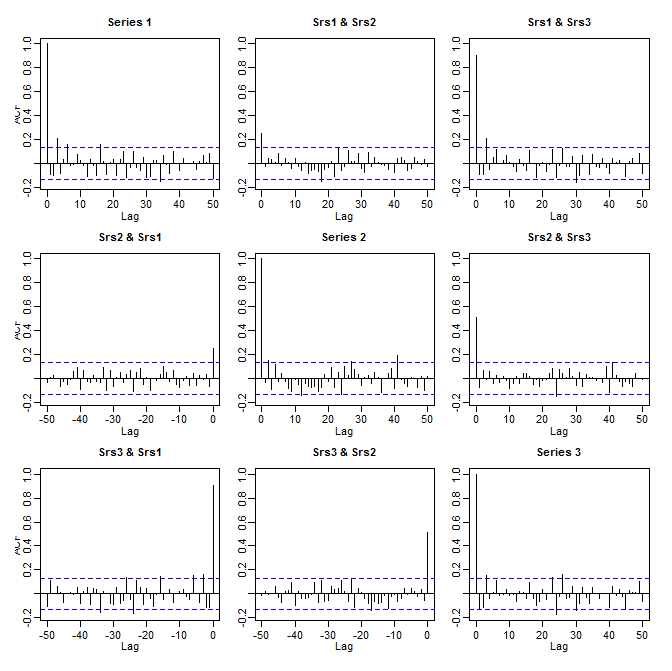}
\includegraphics[width = 3 in]{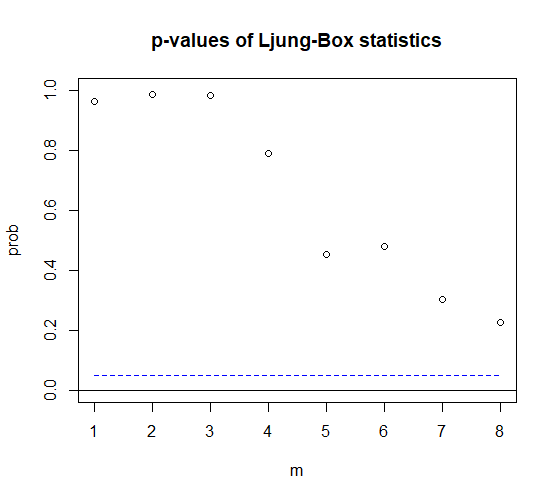}
\caption{Auto-correlation and cross-correlation plot for the estimated residual series (left) and Ljung-Box p-values based on the estimated residuals (right).}
\label{fig:GDP_box_ljung}
\end{figure}

The diagnostics checks based on residual analysis  for the ML fit reveal that the residuals are well-behaved,   and they  mimic the properties of standard multivariate white noise residuals. Figure~\ref{fig:GDP_box_ljung} shows the cross-correlation plot for the residuals.   The auto- and cross-correlations at higher order lags are close to zero. Also Figure~\ref{fig:GDP_box_ljung} shows
 the  $p$-values at different lags for a multivariate Ljung-Box test based on the residuals.  

The estimated long-run equilibrium matrix ${\widehat{\Pi}}_{MLE}$ is 
\[ 
{\widehat{\Pi}}_{MLE} = \begin{bmatrix}  -0.023 &0.024 & 0.028 \\0.022 &-0.028&  -0.016\\ -0.007& 0.023&  -0.028 \end{bmatrix}, 
\]
and the estimated difference operator is $(I_3 - UB)$, where 
\[
 U = \begin{bmatrix} 0.674 &0.431 &0.185\\ 0.430 &0.275 &0.118\\ 0.185& 0.118& 0.051 \end{bmatrix}.
\]

\begin{figure}
\centering
\includegraphics[width = 3 in]{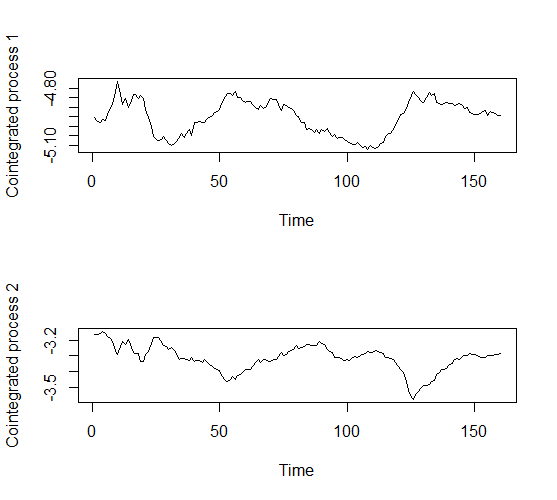}
\caption{Co-integrated processes: co-integrating vectors  ${\beta}_1 = (0.560, -0.665,  -0.492)^{\prime}$ and ${\beta}_2 = (-0.108, 0.530,  -0.841)^{\prime}$.}
\label{fig:coint_GDP}
\end{figure}

Based on the estimated ${\widehat{\Pi}}_{MLE}$, the estimated  co-integrating vectors  are ${\beta}_1 = (0.560, -0.665,  -0.492)^{\prime}$ and ${\beta}_2 = (-0.108, 0.530,  -0.841)^{\prime}$.   A plot of the 
co-integrated processes $\beta^{\prime} Y_t$ is shown in Figure~\ref{fig:coint_GDP}. One of the  processes clearly exhibits nearly nonstationary behavior.  
When an AR(1) is fit to it, the  coefficient estimate is $0.961$, which explains the nonstationary  features of the estimated co-integrated series. 

\begin{figure}
\centering
\includegraphics[height = 6in, width = 3 in]{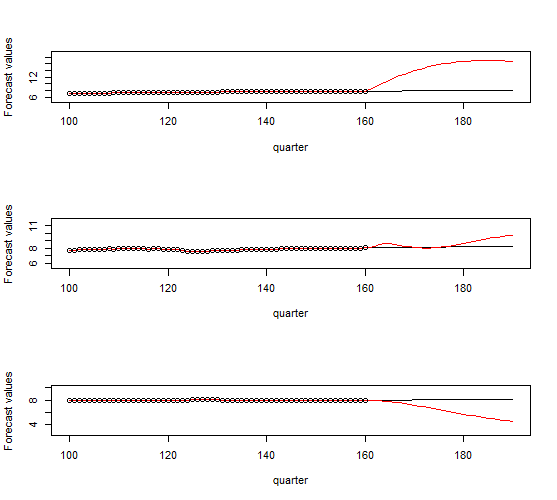}
\caption{The Gross Domestic Product, Import and Export series (2004 Q1 - 2018 Q1 marked in black dots); 
forecast values for the next thirty quarters: OLS (red) and Constrained (black).}
\label{fig:forecast}
\end{figure}

To draw a comparison between estimation with and without the constraints, we studied the behavior of long horizon forecasts
 for the unconstrained OLS (MLE) and the constrained MLE. Specifically, Figure~\ref{fig:forecast} shows one through thirty step ahead forecasts
 for the three series using both the proposed constrained estimator and the unconstrained OLS estimator. The OLS procedure 
 estimates roots that are in the stationary region, but close to unity. As such the roots are similar to that of the constrained estimator,
 but the estimates of constant term are very different for the two methods. The OLS forecast shows explosive behavior in the short term but reverts back to the mean in the
 long term forecast. The proposed constrained estimator provides more reasonable forecasts. 

\section{Discussion}
In this paper we have proposed a paradigm for analyzing VAR models with some unit roots,   some stable roots and some roots exactly equal to zero. Such a framework is based on the idea of exact differencing,   which means factorizing the VAR matrix polynomials to factors involving only unit root and only stable roots.   A parametrization of such factors in terms of unconstrained quantities is obtained.   Such a parameterization makes it feasible to estimate the process with the parameter restrictions  imposed on the estimated VAR polynomial.   The other advantage is that the proposed parameterization can be potentially useful for describing lower dimensional  models in larger dimensional VAR systems.   

While we have restricted the current investigation to the co-integrated model (real unit root equal to one),  the proposed methodology may be extended to seasonally co-integrated processes (with complex unit roots of magnitude one).  Possible factorizations  of first order seasonally co-integrated processes were
 briefly discussed in the  Section 2, on the  commuting factorization of VAR($1$) co-integrated processes.   Such factorizations for general higher order seasonally co-integrated processes in terms of difference   matrices that are periodic of appropriate seasonal period will be investigated in the future.

\Appendix

\renewcommand{\theequation}{A.\arabic{equation}}
\renewcommand{\thetable}{A.\arabic{table}}

\setcounter{equation}{0}

\setcounter{table}{0}

\section{Proofs and additional remarks}

We could assume that $C_{\Phi}$ is semi-simple,
 since the set of polynomials with diagonalizable companion matrices are dense in ${\bar{\mathfrak{S}}}^{m,k}$. But for the present application it suffices to assume that the unit roots and the zero roots are regular. Assuming that the unit roots are regular  rules out processes such as 
\[ 
y_t = \begin{pmatrix} 1 & 0 & 0\\ 1 & 1 & 0 \\ 0 & 0 & 0.5\end{pmatrix} y_{t-1} + \epsilon_t,
 \]
where in spite of  being a  three dimensional VAR($1$)  process with two unit roots, the first difference of the process is not stationary. 
A consequence of the assumption is that $\mbox{rank} (\Pi) = m - r$, where $\Pi = \sum_{j=1}^p \Phi_j - I_m.$

\vspace{1cm}

\noindent {\bf Proof of Proposition \ref{prop:commuting-decomp}.}
Let 
\[
 \Phi = P \, \begin{pmatrix} J_1 & 0 \\ 0 & {\bar{J}} \end{pmatrix} \,  P^{-1}
\]
   be the real Jordan Canonical Form for $\Phi$. Then let 
\[
  {\bar{\Upsilon}} =  P \, \begin{pmatrix} 0 & 0 \\ 0 & {\bar{J}} \end{pmatrix} \, P^{-1}
 \qquad U =  P \, \begin{pmatrix} J_1 & 0 \\ 0 & 0 \end{pmatrix} \, P^{-1}.
\]
 Hence, it follows that
\begin{align*}
  \Phi (B) & = P \,  \left[ I_m -  \begin{pmatrix} J_1 & 0 \\ 0 & {\bar{J}} \end{pmatrix} \, B \right] \,   P^{-1} \\
 \Delta_U (B) & = P \,  \left[ I_m -  \begin{pmatrix} J_1 & 0 \\ 0 &  0  \end{pmatrix} \, B \right] \,   P^{-1} \\
 I_m - \bar{\Upsilon} \, B & = P \,  \left[ I_m -  \begin{pmatrix} 0 & 0 \\ 0 & {\bar{J}} \end{pmatrix} \, B \right] \,   P^{-1}.
\end{align*}
As a result,
\[
  \left(  I_m - \bar{\Upsilon} \, B  \right) \, \Delta_U (B) = \Phi (B) = \Delta_U (B) \, \left(  I_m - \bar{\Upsilon} \, B  \right),
\]
 demonstrating that $({\bar{\Upsilon}}, \bar{U}) \in \mathcal{C}^L_{\Phi} \bigcap \mathcal{C}^R_{\Phi}.$ 
 Now let $( \underline{\Upsilon}, \underline{U})$ be any other pair in $\mathcal{C}^L_{\Phi} \bigcap \mathcal{C}^R_{\Phi}.$ 
 Note that for any positive integer $k \geq 1,$ 
\[  
\Phi^k = {\bar{\Upsilon}}^k + \bar{U} 
\]
 due to $\bar{\Upsilon} \, \bar{U} = 0$ and the idempotency of $\bar{U}$; the same is true of the other pair, i.e.,
\[  
\Phi^k = {\underline{\Upsilon}}^k + \underline{U}. 
\]
 Thus,  for any $k \geq 1$
\begin{equation}
\bar{U} - \underline{U} = \underline{\Upsilon}^k - {\bar{\Upsilon}}^k.
\label{eq:U-difference}
\end{equation}
 Because both ${\bar{\Upsilon}}$ and $\underline{\Upsilon}$ have spectral radius strictly less than one, taking the limit as $k \tends \infty$ in 
  \eqref{eq:U-difference} yields ${\bar{U}} = {\underline{U}}.$  Hence, $({\bar{\Upsilon}}, \bar{U}) = (\underline{\Upsilon}, {\underline{U}}).$  
  This proves the uniqueness of the pair. 
From the facts that for any nonsingular $Q$ with $Q \, \Phi = \Phi \, Q,$ the pair $(Q \, {\bar{\Upsilon}} \, Q^{-1}, Q \, {\bar{U}} \, Q^{-1})$  belongs to 
$\mathcal{C}^L_{\Phi} \bigcap \mathcal{C}^R_{\Phi}$, and that $\mathcal{C}^L_{\Phi} \bigcap \mathcal{C}^R_{\Phi}$ is non-empty, the result follows. 
$\quad \Box$

\vspace{1cm}

\noindent {\bf Proof of Proposition  \ref{prop:schur-redrank-var1}.}
If $\Upsilon = V_1 \, Q_1  \, (I_m + V_1 \, V_1^{\prime})^{-1/2}$, 
  then it is of rank $r$ and satisfies the Riccati equations \eqref{eq:riccati}. Hence it is Schur-stable and of rank $r$.  
 Now suppose $\Upsilon$ is Schur-stable of rank $r$.
    Then let $\Upsilon = \alpha \, \beta^{\prime}$ be a full rank factorization, where $\alpha, \beta$ are $m\times r$ matrices.
   Obviously, the pair$(\alpha, \beta)$ is not unique since any pair of the form $(\tilde{\alpha}, \tilde{\beta}) = (\alpha \, R, \beta \, R^{\dagger})$ 
  will be another full rank factorization for any nonsingular $r\times r$ matrix $R$, where the notation $A^{\dagger}$ stands for the inverse transpose of $A$. 
We could choose $R$ such that ${\tilde{\beta}}^{\prime} \, (I_m + V)^{-1} \, {\tilde{\beta}} = I_r.$ 
  Then for some orthogonal $Q$,  we have ${\tilde{\beta}}^{\prime}  = Q \, (I_m + V)^{-1/2}.$  Also,
   since $\alpha$ is arbitrary, we can choose $\alpha$ such that  ${\tilde{\alpha}} \, {\tilde{\alpha}}^{\prime} =V.$ 
  Thus, choose ${\tilde{\alpha}} = V_1$ as a lower triangular $m\times r$ matrix will work. 
$\quad \Box$

\vspace{1cm}

\noindent {\bf Proof of Proposition \ref{prop:reduced_last_coeff}.}
 The companion matrix is 
\[
 C_A = \left[ \begin{array}{lllll}  A_1 & A_2 & \ldots & A_{k-1} &  A_k \\	
					I_m & 0 & \ldots & 0 & 0 \\
					\vdots & \vdots & \vdots & \vdots & \vdots \\
					0 & 0 & \ldots & I_m & 0 
		\end{array} \right].
\]
  We can block-column permute $C_A$, without changing the determinant, by shifting the last block column to the first block column.
 The resulting matrix is block upper-triangular, with determinant equal to $\det A_k$, i.e., $\det C_A = \det A_k$.  $\quad \Box$

\vspace{1cm}

\noindent {\bf Proof of Theorem  \ref{thm:varp}.}
Given  $\Phi(z) \in {\bar{\mathfrak{S}}}^{m,k}_{r,0}$, let $U  = P_{\mathcal{N}(\Pi)}$ be the projection matrix
 for the null space $\mathcal{N}(\Pi)$, where $\Pi = \sum_{i=1}^k \Phi_i - I_m.$ 
Then $U$ satisfies the conditions of the theorem. 
Define $\Upsilon(z)  = (I_m - \sum_{j=1}^k \Upsilon_j z^j)  \in {\mathfrak{S}}^{m,k}$ by 
\begin{align}
\Upsilon_1 &= \Phi_1 - U \nonumber \\
\Upsilon_j &= \Phi_j + \sum_{i=1}^{j-1} \Phi_i \, U - U, \; j = 2, \ldots, k. 
\label{eq:Upsilon_from_phi}
\end{align}
Then $\Upsilon_k \, U = \Pi \, U = 0,$ and $\Phi(z) = \Upsilon(z) \, \Delta_U(z).$
Also because $\Upsilon_k \, U = 0$, $ \mbox{dim} (\mathcal{N}(\Upsilon_k)) \geq r.$ 
 Next, we show this dimension exactly equals $r$.
 Assume there exists a vector $x$ such that $\Upsilon_k \, x = 0$ and $x \notin \mathcal{C}(U),$ the column space of $U$. 
From \eqref{eq:Upsilon_from_phi}, we have 
\begin{equation}
 \label{eq:x.nullvec.ups}
 0 = \Upsilon_k \, x = \Pi \, U  \, x + \Phi_k  \, (I_m - U) \, x = \Phi_k \, (I_m - U) \, x.
\end{equation}
Because $\Phi(z) \in {\bar{\mathfrak{S}}}^{m,k}_{r,0}$, $\mbox{rank} (\Phi_k) = m$ and hence it is invertible. 
 Premultiplying (\ref{eq:x.nullvec.ups}) by the inverse of $\Phi_k$, 
we have  $x = U \, x$,  leading to a contradiction. Thus, $\mathcal{N}(\Upsilon_k) = \mathcal{C}(U),$ and $\Upsilon(z)  \in {\mathfrak{S}}^{m,k}_r$. 
 
To show uniqueness of the choice of $(\Upsilon(z), U)$, suppose there is another pair $(\Upsilon^*(z), U^*)$ satisfying (\ref{eq:leftfact}).   Then 
\[ 
\Pi \, U^* = \Upsilon_k^* \,  U^* = 0,
\]
where the first equality follows because the coefficients of $\Upsilon^*(z)$ must satisfy \eqref{eq:Upsilon_from_phi} and 
 the second equality follows because for any $(\Upsilon, U)$ satisfying  $\Phi(z) = \Upsilon(z)\Delta_U(z)$ for all $z \in \mathbb{C},$ 
 we must have the coefficient of $z^{k+1}$ on the right hand side, 
 $\Upsilon_k U$, equal to zero.  Then by assumption, $\mathcal{C}(U^*) = \mathcal{N}(\Pi). $
 Because $U^*$ is symmetric and idempotent it implies $U^* = P_{\mathcal{N}(\Pi)}$ is the unique projection matrix for  $\mathcal{N}(\Pi).$ 
Then from the relation \eqref{eq:Upsilon_from_phi}  we have $U^* = U$. 

Next we show that the nonzero roots of $\det {\tilde{\Upsilon}}(z) = 0$ are the same as the stable roots of 
 $\det {\tilde{\Phi}}(z) = 0.$  Suppose $(\lambda, x)$ is a pair of eigenvalue and eigenvector
 for the companion matrix of $\Phi$ and let $0 < |\lambda| < 1.$  Then $x$ must be of the form 
$x^{\prime} = (x_1^{\prime}, \lambda^{-1}x_1^{\prime}, \ldots, \lambda^{k-1} x_1^{\prime})$ for some $m\times 1$ vector $x_1$. Also 
\begin{equation}
 \sum_{j=1}^k \lambda^{-(j-1)}  \Phi_j x_1 = \lambda x_1.
\label{eq:eigen_companion}
\end{equation}
 But following the identity (\ref{eq:leftfact}), the coefficients of $\Phi$ can be written as 
\begin{eqnarray}
\Phi_1 &=& \Upsilon_1  +  U \nonumber \\
\Phi_j &=& \Upsilon_j - \Upsilon_{j-1} U, \; j = 2, \ldots, k. 
\label{eq:phi_from_Upsilon}
\end{eqnarray}
Thus, from \eqref{eq:phi_from_Upsilon} we have that  \eqref{eq:eigen_companion} holds if and only if
\begin{align*}
& (\Upsilon_1 + U) \, x_1 +  \sum_{j=2}^k \lambda^{-(j-1)} \, [\Upsilon_j - \Upsilon_{j-1} U] \, x_1 = \lambda x_1 \\
& \Leftrightarrow   \sum_{j=1}^k \lambda^{-(j-1)} \Upsilon_j \, [I_m - \lambda^{-1} U] \, x_1 = \lambda [I_m - \lambda^{-1} U] \, x_1,
\end{align*}
where the last equality is obtained using the fact that $\Upsilon_k \, U = 0$. 
Thus, $(\lambda, (I_m - \lambda^{-1}U) \, x_1 )$ is a pair of eigenvalue and eigenvector 
 for the companion matrix of $\Upsilon(z).$ Hence, the stable roots of the VAR polynomial
 $\Phi(z)$ are the same as the nonzero roots of $\det {\tilde{\Upsilon}}(z) = 0.$ 

For the converse, let $\Upsilon(z) \in {\mathfrak{S}}^{m,k}_r$ be given.
 Define $U = P_{\mathcal{N}(\Upsilon_k)}$, the projection matrix of the null space of $\Upsilon_k.$ 
Define $\Phi$ following \eqref{eq:phi_from_Upsilon}. Then $U$ is symmetric and idempotent of rank $r$. 
Also, writing $\Pi$ in terms of $\Upsilon$ it is easy to see that $\Pi \, U = 0$ and hence $\mathcal{C}(U)$ is a subspace of $\mathcal{N}(\Pi)$,
  where $\Pi = \sum_{j=1}^k \Phi_j - I_m.$ Now let $x \in \mathcal{N}(\Pi)$ but $x \notin \mathcal{C}(U)$. From $ 0  = \Pi \, x $ 
 and using 
\[
 \Pi = \sum_{j=1}^k \Upsilon_j - \sum_{k=1}^{j-1} \Upsilon_j \, U + U - I_m,
\]
we have 
\[ 
[I_m - \sum_{j=1}^k \Upsilon_j] \, x = [I_m -  \sum_{j=1}^k \Upsilon_j] \, U \, x. 
\]
Since $\Upsilon(z) \in {\mathfrak{S}}^{m,k}_r$ by assumption, $[I_m -  \sum_{j=1}^k \Upsilon_j] $ is invertible.
 Hence $x = U \, x$, a contradiction; therefore   ${\mathcal{N}(\Upsilon_k)} = {\mathcal{N}(\Pi)}$ and
  $U = P_{\mathcal{N}(\Upsilon_k)} = P_{\mathcal{N}(\Pi)}.$ Thus, $\det{\tilde{\Phi}}(z) = 0$
 has exactly $r$ unit roots and $\Phi(z) \in {\bar{\mathfrak{S}}}^{m,k}_{r}.$ Also, following similar arguments as in the first part of the proof, 
  we see that the stable  roots of $\det {\tilde{\Phi}}(z) = 0$ are the same as the nonzero roots of $\det {\tilde{\Upsilon}}(z) = 0.$
Thus, $\Phi(z) \in {\bar{\mathfrak{S}}}^{m,k}_{r,0}.$ To show uniqueness, one can proceed exactly same way as described above. 
 $\quad \Box$

\vspace{1cm}

\noindent {\bf Proof of Theorem  \ref{thm:factor-leftright}.}
From the proof of Theorem~\ref{thm:varp}, it is easily seen that a difference matrix $U$ in the left factorization
 is a projector (not necessarily the orthogonal projector) to $\mathcal{N}(\Pi)$,
 and similarly each $U$ in the right factorization is a projector to  $\mathcal{N}(\Pi^{\prime})$.  
Any matrix   $P$ is a projector to $\mathcal{N}(\Pi)$ if and only if $I_m - P$ is a projector to $\mathcal{C}(\Pi)$,
   and any ${\tilde{P}}$ is a projector  to $\mathcal{N}(\Pi^{\prime})$ if and only if $I_m - {\tilde{P}}$ is a projector
  to $\mathcal{C}(\Pi^{\prime})$.   Establishing a bijection between the projectors of the  column space and the row space completes the proof. 
 Note that the projector to $\mathcal{C}(\Pi)$ will be of the form $P = \Pi \, B$ for some square matrix $B$,
    which will be a generalized inverse of $\Pi$. Then we could construct ${\tilde{P}}$, a projector to $\mathcal{N}(\Pi^{\prime})$ as  ${\tilde{P}} = \Pi^{\prime} \, \tilde{B}$,
  where $\tilde{B}$ is the associated generalized inverse of $ \Pi^{\prime}.$   This provides the one-to-one mapping between the projectors.  
 Once the bijection between the differencing matrices $U$ has been established for the left and right factorizations, the remaining coefficients
  can be mapped using the relation between $\Phi$ and $\Upsilon$ for a given $U$ for the left and the right factorization. 
$\quad \Box$

\vspace{1cm}

\noindent {\bf Proof of Theorem   \ref{thm:schur_iter}.}
The first part
 follows from Theorem~3 in Roy et al. (2019). From the Yule-Walker form of the polynomial, we have 
\[ 
 [\Upsilon_1^{\prime}, \ldots, \Upsilon_k^{\prime}] = [U(1), \ldots, U(k)] \, {\underline{U}}_{k-1}^{-1}.
\]
Using the formula for inverse of partitioned matrices, we have 
\begin{eqnarray*}
\Upsilon_k^{\prime} &=&  [\xi_{k-1}^{\prime}, \; U(k)]\begin{pmatrix} -{\underline{U}}_{k-2}^{-1}\kappa_{k-1}D_{k-1}^{-1} \\ D_{k-1}^{-1} \end{pmatrix} \\
&=&(U(k) - \xi_{k-1}^{\prime}{\underline{U}}_{k-2}^{-1}\kappa_{k-1}) \, D_{k-1}^{-1}.
\end{eqnarray*}
 Using this fact, together with calculations from Theorem~3 of Roy et al. (2019),  we have 
\begin{eqnarray}
C_{k-1} - C_k &=& (U(k) - \xi_{k-1}^{\prime}{\underline{U}}_{k-2}^{-1}\kappa_{k-1}) \, D_{k-1}^{-1} \, (U(k) - \xi_{k-1}^{\prime}{\underline{U}}_{k-2}^{-1}\kappa_{k-1})^{\prime}\nonumber \\
 &=& \Upsilon_k^{\prime} \, D_{k-1} \, \Upsilon_k.
\label{eq:rank_equal}
\end{eqnarray}
Since ${\underline{U}}_k$  is a positive definite matrix, all principal minor and the Schur complements are positive definite.
  Thus, $D_{k-1}$ is a positive defnite matrix. Hence from \eqref{eq:rank_equal}, it follows that 
\[
 \mbox{rank} (\Upsilon_k) = \mbox{rank} (C_{k-1} - C_k).  \quad \Box
\]

\vspace{1cm}

\noindent {\bf Proof of Proposition \ref{prop:ortho-matrix}.}
If $A \, Q = B$ then it is obvious that $B \,  B^{\prime} = A \, Q \, Q^{\prime} \, A^{\prime}  = A \, A^{\prime}.$ 
 Now suppose  $A \, A^{\prime} = B \, B^{\prime}$. Since $A$ has full column rank, there exists a $p\times m$ matrix 
 $C$ such that $C \, A = I_p$. Thus, $I_p = C \, B \, B^{\prime} \, C^{\prime}.$  Hence choosing $Q = C \, B$ makes $Q \, Q^{\prime} = I_p.$
  Since $A$ and $B$ have the same column space, there is a $p\times m$ matrix $D$ such that $B = A \, D.$ This implies 
\[
 A \, Q = A \, C \, B = A \, C \, A \, D = A \, D = B. \quad \Box
 \]

\end{document}